\documentstyle[espcrc2,fleqn,twoside,epsfig]{article}

\title{Theory of the Optical Properties of 
a DNA-Modified Gold Nanoparticle System}
\author{Sung Yong Park and David Stroud
\address{Department of Physics,
The Ohio State University, Columbus, Ohio 43210}
}

\begin{document}

\def\etal{ {\it et.\ al.}}
\def\eps{\epsilon}
\def\tnsr{\tensor}
\def\beq{\begin{equation}}
\def\eeq{\end{equation}}


\begin{abstract}
\vspace{2pc}

We describe a simple model for the melting and optical properties of a DNA/gold nanoparticle aggregate.
The aggregate is modeled as a cluster of gold nanoparticles on a periodic lattice connected by DNA bonds,
and the extinction coefficient is computed using the discrete dipole approximation. 
The optical properties at fixed wavelength change dramatically at the melting transition, 
which is found to be higher and narrower in temperature for larger particles, and much sharper than 
that of an isolated DNA link.  All these features are in agreement with available experiments.

\vspace{5pc}
\end{abstract}
\maketitle



\section{Introduction}
Recently, so-called functional metallic nanoparticles have started to be developed, 
which may lead to new materials with improved optical and mechanical properties~\cite{sanchez}.  
Among these, there is a particular interest in gold nanoparticles to which noncomplementary 
oligonucleotides capped with thiol groups are attached (DNA modified gold nanoparticle system), because
in addition to a strategy using self-assembly of nanoparticles~\cite{mirkin}, biological detection is
possible using the optical and electrical sensitivity of their aggregates~\cite{elghanian,sjpark}.
 
In this paper, we present a  model for 
this ``melting'' transition, which accounts for most experimentally observed 
features.

\section{Structural Modeling}

To calculate the $T$-dependent optical properties, we have considered two slightly different structural 
models.  
In the simplest version of our model, the low-$T$ aggregate is 
taken simply as a collection of identical gold nanospheres 
(each of radius $a$) which form a cube of edge $L$ and fill the sites of a simple cubic lattice of lattice constant $d$ ($d > 2a$), 
containing $N_{\rm par} = (L/d)^3$ gold nanoparticles. 
We have also investigated the melting and the optical properties assuming 
that the low-$T$ cluster is a fractal aggregate.
In both cases, we assume that all bonds are occupied by the same number ($N_s/z$) of DNA links at low temperature,
where $N_s$ is the number of single DNA strands attached 
to one particle and $z$ is the number of nearest neighbors
($z = 6$ for a simple cubic lattice).  

\begin{figure}
\epsfxsize=8cm \epsfysize=6cm \epsfbox{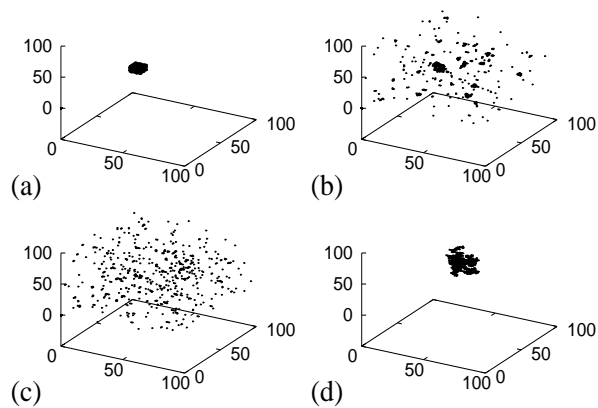} 
\caption{Schematic of the melting of a gold-DNA cluster, for
two different models discussed in the present paper.  In
the first model, (a) shows the cluster at low $T$ ($p_{\rm eff}=1$);
(b) $p_{\rm eff}=0.5 > p_c(L)$, 
where $p_c$ is the weakly $L$-dependent percolation threshold;
(c) $p_{\rm eff}(T) = 0.2 < p_c(L)$;
(d) alternate model for low-$T$ sample ($p_{\rm eff}(T) = 1$):
fractal cluster formed by
reaction-limited cluster-cluster aggregation [RLCA],  
with fractal dimension $d_f \sim 2.1$.}
\label{fig:1}
\end{figure}

As temperature is increased, the bonds start to break.
We define the probability that a link forms between a given gold particle pair at temperature $T$ as $p_{\rm eff}(T)$.
To generate a specific sample at temperature $T$, we randomly remove links with probability $1 - p_{\rm eff}(T)$, 
then identify the separate clusters, using a simple computer algorithm~\cite{stauffer}.  
If the aggregate is separated into two or more clusters, we simply place these clusters in random positions and orientations 
within a larger bounding box (usually of edge 100$d$), taking care that the individual clusters do not overlap.   
The resulting geometry is shown schematically in Fig.\ \ref{fig:1}.  As a second structural model, we have carried out the same procedure 
to simulate the melting of a sample formed by reaction-limited cluster-cluster aggregation 
(RLCA)~\cite{brown}.  A typical RLCA cluster is shown in Fig.\ \ref{fig:1} (d), and represents a
possible fractal aggregate which might be produced by certain random growth processes at low temperature~\cite{sypark}.  
Here we implicitly assume that the specific bonds which are occupied at temperature $T$ are time-independent.

\section{Melting Theory for DNA Links}

The two-state model can describe melting well for short DNA strands (12-14 base pairs)~\cite{bloomfield,werntges},
and is  given by the relation
\begin{equation}
S + S \rightleftharpoons D.
\label{eq:react}
\end{equation}
The chemical equilibrium condition corresponding to (\ref{eq:react}) is
\begin{equation}
\frac{[1-p(T)]^2}{p(T)} = \frac{K(T)}{C_T}, 
\label{eq:equil}
\end{equation}
where $p(T)$ is a (temperature-dependent) fraction of the double DNA strands,
$K(T)$ is a suitable chemical equilibrium constant, 
and $C_T$ is the molar concentration of single DNA strands in the sample.
We have also assumed the simple van't Hoff behavior 
$K(T) = \exp[-\Delta G/k_BT]$, with a Gibbs free energy of formation
$\Delta G(T) = c_1(T - T_M) + c_2(T - T_M)^3$, choosing the values of 
$c_1$, $c_2$, and $T_M$ to be consistent with experiments on these DNA molecules.

In order to calculate the fraction $p_{\rm eff}(T)$ of {\em bonds} which
contain at least one double strand, we adopt the following very 
simplified model.  First, we assume that exactly $N_s/z$
of the single strands on a given nanoparticle are available
to bond with any {\em one} of its $z$ nearest neighbors.  
Thus, we assume that the maximum number of links that could be 
formed between any two particles is $N_s/z$. The probability that
{\em no} link is formed is then taken to be
\begin{equation}
1 - p_{\rm eff}(T) = [1 - p(T)]^{N_s/z}.
\end{equation}

\begin{figure}
\epsfxsize=8cm \epsfysize=6cm \epsfbox{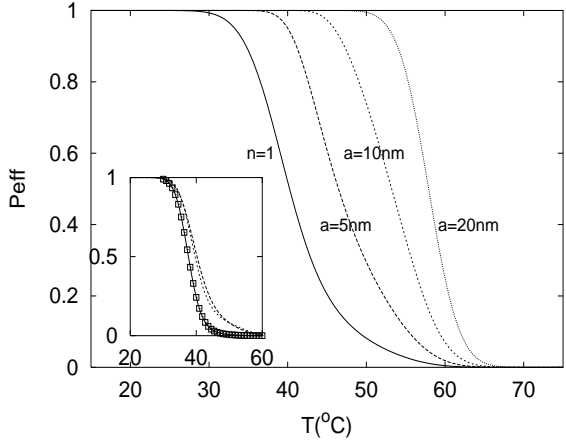} 
\caption{Plot of $p_{\rm eff}(T)$ versus $T$ for several different
choices of particle radius $a$, as indicated.
Also plotted is $p(T)$, the probability that a given DNA strand 
is part of a double strand at $T$, (indicated as $n=1$).
Inset: Comparison of $p(T)$ and $p'(T)$.}
\label{fig:2}
\end{figure}

In Fig.\ \ref{fig:2}, we plot p$_{\rm eff}$(T) for several radius $a$.  We
assume $N_s \propto a^2$, set $z = 6$ and use the experimental result 
that $N_s = 160$ when $a = 8$ nm~\cite{demers}.  

In actuality, there is a linker molecule which emerges from
solution to connect two DNA single strands on different 
nanoparticles.  But in fact, we can avoid considering the
linker molecule explicitly, as shown in the inset of Fig.\
\ref{fig:2}.   In this inset, 
the dotted and dashed lines represent $p(T)$ for
two melting curves, one for each single strand connected to a
linker molecule with slightly different constants $c_1$ and $c_2$
but the same melting temperature.  (This is the case where
the linker molecule is expected to have the most effect.)  
The solid line is the
resulting melting curve for the two single strands plus linker.
The squares are the melting probability $p^\prime(T)$ calculated
assuming two effective strands and no linker strand with slightly 
different $c_1$, $c_2$, and $T_M$.

\section{Optical Properties}

We calculate the optical properties of this sample using the Discrete Dipole Approximation (DDA)~\cite{pp,draine}.  
The sample is modeled as a collection of separate aggregates, whose extinction coefficients are computed individually,
then added.  Each aggregate consists of many identical nanoparticles, which have complex frequency-dependent 
dielectric constant $\epsilon(\omega)$, and polarizability $\alpha(\omega)$ related to $\epsilon(\omega)$
by the Clausius-Mossotti equation with radiative reaction correction term~\cite{draine1}.
The resulting expressions for the induced dipole moment ${\bf p}_i$ of the $i^{th}$ sphere, and the corresponding expression
for the extinction coefficient $C_{\rm ext}(k)$ at wave number $k$, are given in Ref.\ \cite{lazar}.

In our case, each cluster consists of a number of DNA-linked individual gold nanoparticles.
In our calculations, we do not include the optical properties of the DNA molecules, since
these absorb primarily in the ultraviolet~\cite{storhoff1}.  
We use tabulated values of the gold complex index of refraction~\cite{lynch,johnson}, then  calculate $C_{\rm ext}$ for each cluster 
using the DDA.  To improve the statistics, we average $C_{\rm ext}$ for each cluster over possible orientations.  
We then sum the averaged extinction coefficients of all the individual clusters to get the total extinction coefficient 
of the suspension.
This method is justified when the suspension is dilute.

\begin{figure} 
\epsfxsize=8cm \epsfysize=6cm \epsfbox{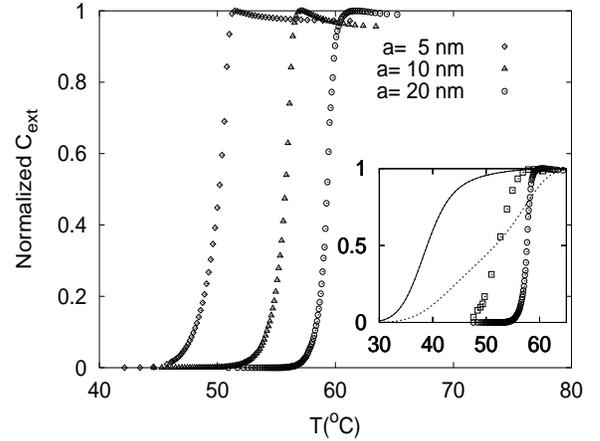} 
\caption{Normalized extinction coefficient $C_{\rm ext}(\lambda, T)$
for $\lambda = 520$ nm, plotted versus $T$ 
for several particle radii $a$, assuming
that the low-$T$ aggregate is an $N = 1000$ simple cubic
cluster, as shown in Fig.\ \ref{fig:1}. 
Inset: Normalized extinction coefficient $C_{\rm ext}(\lambda, T)$,
versus $T$ for $\lambda = 520$ nm, plotted for a 1000-particle
gold/DNA aggregate assuming that the low-temperature
sample is a simple cubic cluster (open circles) or
an RLCA cluster (open squares). The solid curve is a plot
of $1 - p(T)$ for a single DNA duplex with the same
concentration $C_T$ as the above two curves.
The dotted curve represents $1 - p(T)$ for a 
single DNA duplex but with a much higher $C_T$ than for
the other curves of the inset.}

\label{fig:3}
\end{figure}

In Fig.\ \ref{fig:3}, we show $C_{\rm ext}(\lambda, T)$ at {\em fixed}
$\lambda = 520$ nm, close to the isolated-particle 
surface-plasmon resonance (SPR), versus $T$, for several particle sizes, 
assuming that the low-T aggregate is a simple cubic cluster with $N = 1000$ 
particles, which is illustrated in Fig.\ \ref{fig:1} (a). 
As can be seen in Fig.\ \ref{fig:3}, for each radius $a$, 
the extinction increases sharply at a characteristic temperature, 
corresponding to the melting of the aggregate for that size; 
at this temperature, the absorption due to the SPR increases sharply.   

In the inset of Fig.\ \ref{fig:3}, we compare 
$C_{\rm ext}(\lambda, T)$ for a regular and an 
RLCA cluster of particles of $20$ nm radius at 
$\lambda$ = 520 nm, both for $N = 1000$ 
particles.  Although the RLCA cluster has a 
slightly broader melting transition, as 
manifested in $C_{\rm ext}(\lambda, T)$, than does the regular lattice,
both sets of data show a much sharper melting transition than
that of a single DNA link.  Also, although our normalized
$C_{\rm ext}(\lambda, T)$ is calculated for the aggregates
at 520 nm, we expect similar behavior at $260$ nm. 
(We have not carried out calculations at this $\lambda$ mainly 
because we have not included the DNA absorption properties.)  
In any case, the experimental melting curves at 260 and 520 nm 
are very similar~\cite{kiang}. 

\section{Discussion}

We have briefly described a model for structural development
of DNA/gold nanoparticle clusters, and have calculated the 
resulting cluster optical properties using the DDA.  

Our calculated extinction coefficients are good agreement with 
recent experimental results~\cite{elghanian,storhoff1,kiang}.
In particular, both experiment and calculation give a sharp increase 
in $C_{\rm ext}(\lambda, T)$ at fixed $\lambda$, as $T$ increases 
past a critical temperature. 
We also find, in agreement with experiment, that melting occurs over 
a much narrower range of $T$ in the aggregate than for a single bond, 
and that the melting occurs at higher $T$ for larger particles.

\section{Acknowledgments.} This work has been supported by NSF Grant
DMR01-04987, by the U.-S./Israel Binational Science Foundation, 
and by an Ohio State University Postdoctoral Fellowship.  
Calculations were carried out using the facilities of 
the Ohio Supercomputer Center.

\end{document}